\documentclass[twocolumn,preprintnumbers,amsmath,amssymb]{revtex4-2}
\usepackage{amsmath}
\usepackage{amssymb}
\usepackage{graphicx}
\usepackage{bm}
\usepackage{mathrsfs}
\usepackage{subfigure}
\usepackage[breaklinks=true,colorlinks,citecolor=blue,linkcolor=blue,urlcolor=blue]{hyperref}

\begin{document}
\title{Optical absorption window in Na$_3$Bi based three-dimensional Dirac electronic system}

\author{Q. N. Li$^{1}$}
\author{W. Xu$^{1,2}$}\email{wenxu$_$issp@aliyun.com}
\author{Y. M. Xiao$^{1}$}\email{yiming.xiao@foxmail.com}
\author{L. Ding$^{1}$}
\author{B. Van Duppen$^{3}$}
\author{F. M. Peeters$^{1,3}$}

\address{$^1$ School of Physics and Astronomy and Yunnan Key Laboratory for Quantum Information,
Yunnan University, Kunming 650504, China}
\address{$^2$ Key Laboratory of Materials Physics, Institute of Solid State Physics, Chinese Academy
of Sciences, Hefei 230031, China}
\address{$^3$ Department of Physics, University of Antwerp, Groenenborgerlaan 171, B-2020 Antwerpen, Belgium}

\date{\today}

\begin{abstract}
We present a detailed theoretical study of the optoelectronic properties of a
Na$_3$Bi-based three-dimensional Dirac electronic system (3DDES). The optical
conductivity is evaluated using the energy-balance equation derived from a
Boltzmann equation, where the electron Hamiltonian
is taken from a simplified $\mathbf{k}\cdotp \mathbf{p}$ approach. We find that
for short-wavelength irradiation, the optical absorption in Na$_3$Bi is mainly
due to inter-band electronic transitions. In contrast to the universal optical
conductance observed for graphene, the optical conductivity for Na$_3$Bi based
3DDES depends on the radiation frequency but not on temperature, carrier density
and electronic relaxation time. In the radiation wavelength regime of about
5 $\mu m<\lambda<$ 200 $\mu m$, an optical absorption window is found. This is
similar to what is observed in graphene. The position and width of the absorption
window depend on the direction of the light polarization and sensitively on
temperature, carrier density, and electronic relaxation time. Particularly, we
demonstrate that the inter-band optical absorption channel can be switched on
and off by applying the gate voltage. This implies that similar to graphene,
Na$_3$Bi based 3DDES can also be applied in infrared electro-optical modulators.
Our theoretical findings are helpful in gaining an in-depth understanding of
the basic optoelectronic properties of recently discovered 3DDESs.
\end{abstract}


\maketitle

\section{\label{sec:level1}{Introduction}}
Since the discovery of graphene \cite{No04}, extensive theoretical
and experimental investigations have been carried on Dirac electronic systems \cite{Al09}.
They exhibit unique and important physical properties that are different from conventional
electronic materials such as metals, semiconductors, oxides, etc. In
a two-dimensional (2D) Dirac system such as graphene, the electrons
are massless and have a gapless and linear-like energy dispersion.
These features lead to excellent electronic, transport and
optical properties applicable in advanced electronic
and optoelectronic devices. Akin to 2D Dirac electronic systems
which are normally realized in atomically thin material
structures, three-dimensional (3D) Dirac electronic systems (3DDESs)
have been recently discovered \cite{Yo12}. These
3DDESs are normally bulk-like with topological electronic
structures. At present, the most popularly studied 3DDESs are based
on Cd$_3$As$_2$ \cite{Ne14} and Na$_3$Bi \cite{Liu14} semi-metals. Such
3DDESs are also gapless and have approximately a linear energy dispersion
for the bulk states in conduction and valence bands around the
topologically protected Dirac points \cite{Yo12}. In these
materials, the crystal inversion symmetry or the time-reversal
symmetry is broken so that each Dirac node splits into a pair of
opposite-chirality Weyl nodes and a Berry curvature can be
achieved in the electronic band structure. As a result, many
unique and novel physical phenomena can be observed in 3DDESs, such
as chiral electro- \cite{Ma12} and magneto-transport \cite{Liang18},
magnetoelectric-like plasmons and photons \cite{As14}, Fermi-arc
surface states \cite{Xu15}, chiral-pumping effect \cite{Eb16}, nontrivial
$\pi$ Berry phase of Dirac fermions \cite{Wang16}, 3D quantum Hall
effect \cite{Zhang19}, etc. These interesting and important research
findings demonstrate that the field of 3DDESs is very rich not only
in fundamental physics but also in potential applications of 3DDESs
in advanced electronic and optoelectronic devices.

Up to now, the study of 3DDESs has been mainly focused on their
electronic structure, electric and magneto transport and plasmonic
properties. Relatively less research has been published on its
optoelectronic properties. Experimentally it was demonstrated that
3DDESs can be applied as infrared photodetectors and
ultrafast optical switches \cite{Huang19,Yang19}. Very recently, 3DDESs have
been applied for terahertz (THz) high harmonic generation \cite{Lu18}.
For basic study and device application of 3DDESs in
optoelectronics, it is of great importance to
examine their fundamental optical properties, which
is the prime motivation of the present study.

In 2DDESs such as graphene, interesting and important optoelectronic properties
have been observed and investigated. In particular, it has been
found that graphene is a wide-band optoelectronic material \cite{No05,Wu12,Bao11}
with unique electronic and optical properties. In the short-wavelength
regime (visible to near-infrared), the optical conductivity of
graphene is caused by inter-band electronic transitions and is
universal \cite{Ku08}. The optical absorption channel can be
turned on and off by varying the Fermi energy or electron density
in graphene as can be realized, e.g., by applying a gate voltage \cite{Wang08,Da08}.
In the long-wavelength regime (terahertz bandwidth), the optical
conductivity in graphene is induced by free-carrier optical
absorption and increases with radiation frequency \cite{Fa07}. In the
intermediate-wavelength regime (mid- to far-infrared), an optical
absorption window can be observed \cite{Ku08,Fa07,Ry09}, which is a
consequence of the fact that different photon energies are required for
inter- and intra-band electronic transitions. The width and depth
of this window depend strongly on temperature and the carrier density in
graphene. These important features have been utilized to realize
graphene-based infrared detectors at ambient condition and electro-optical
modulators \cite{Hau04,Mo01}. For 3DDESs, because of the different
electronic band structure, i.e., the 3D nature of electron
motion and the presence of the Berry curvature in the electron energy spectrum,
one would expect that the basic features of the electronic
transition accompanied by the absorption of photons can differ
significantly from 2DDESs.

In this study, we intend to reveal and examine these features and to compare them
with those already observed experimentally for graphene. We focus our attention
on Na$_3$Bi-based 3DDESs. At present, high-quality hexagon
plate-like Na$_3$Bi crystals with large (001) plane surfaces can be
grown from a molten Na flux \cite{Ku15} and the features of 3D Dirac
fermions can be clearly achieved \cite{Liu14,Wang12}. Our goal is to investigate the
basic optoelectronic properties of the semi-metal Na$_3$Bi. Particularly, we
look into the effect of the Berry curvature and the crystal orientation of Na$_3$Bi
on optical response. From our theoretical results, we explore the possible
application of 3DDESs as novel optoelectronic materials and devices.

This paper is organized as follows. The theoretical approaches to
calculate the electronic band structure, the electron-photon
interaction and the optical conductivity in Na$_3$Bi semi-metal are
presented in Section \ref{sec:leve2}. In Section \ref{sec:leve3}, we present and discuss the
numerical results for the optoelectronic properties of Na$_3$Bi
based 3DDES. Our concluding remarks are summarized in
Section \ref{sec:leve4}.

\section{\label{sec:leve2}{THEORETICAL APPROACH}}
\subsection{\label{sec:leve21}{Simplified electronic band structure}}
In this study, we take Na$_3$Bi as an example to study the basic
optoelectronic properties of 3DDESs. Na$_3$Bi is a hexagonal crystal
with normally the P6$_3$/mmc or D$^4_{6h}$ phase \cite{Wang12}. There
are two nonequivalent Na sites noted as Na(1) and Na(2). Na(1) and
Bi can form simple honeycomb lattice layers stacked along the
$c$-axis. The Na(2) atoms are sandwiched between the above-mentioned
lattice layers and connect to the Bi atoms in forming the layers of
honeycomb lattices. Na$_3$Bi has an inverted band structure and its
Fermi surface consists of two isolated Fermi points \cite{Wang12}. Both
time-reversal and inversion symmetries are present in Na$_3$Bi so
that there is fourfold degeneracy at each Fermi point around which
the band dispersion can be linearized. Due to the lattice structure,
asymmetric features of the electronic band structure can be
expected for Na$_3$Bi Dirac fermions. The Hamiltonian for relatively
low-energy electrons in the semi-metal Na$_3$Bi can
be described within a $\mathbf{k}\cdotp\mathbf{p}$ approximation
\cite{Wang12}, which reads
\begin{equation}\label{1}
H(\mathbf{K})=\epsilon(\mathbf{K})\times I +\left(
\begin{array}{cccc}
M_\mathbf{K} & Ak_+ & 0 & B^*_\mathbf{K}\\
Ak_- & -M_\mathbf{K} & B^*_\mathbf{K} & 0\\
0 & B_\mathbf{K} & M_\mathbf{K} & -Ak_-\\
B_\mathbf{K} & 0 & -Ak_+ & -M_\mathbf{K}\\
\end{array}
\right),
\end{equation}
where $\mathbf{K}=(\mathbf{k},k_z)=(k_x,k_y,k_z)$ is the electron wave-vector
or momentum operator, and $I$ is a 4$\times$4 unitary matrix. The $z$-
direction is taken along the stacking direction of the honeycomb
lattice layers formed by Na(1) and Bi, $k_\pm=k_x\pm ik_y$,
$\epsilon(\mathbf{K})=C_0+C_1k_z^2+C_2k^2$,
$M{_\mathbf{K}}=M_0-M_1k_z^2-M_2k^2$, and $C_0,\ C_1,\ C_2,\ M_0,\
M_1,\ M_2$ and $A$ are band parameters \cite{Wang12}. In this
Hamiltonian, $B_\mathbf{K}=B_3k_zk^2_+\sim K^3$ gives a high-order
contribution to the electron motion, which is significant only at
relatively large electron momentum. The corresponding
Schr\"{o}dinger equation can be solved analytically and the
eigenvalues are obtained as
$E_\pm(\mathbf{K})=\epsilon(\mathbf{K})\pm\sqrt{M^2_\mathbf{K}+A^2k^2+|B_\mathbf{K}|^2}$,
where the upper (lower) case refers to conduction (valence) band.
In this electronic system, two Dirac points exist at $k=0$ and $k_z=\pm
k_c=\pm\sqrt{M_0/M_1}$.

Now, we consider the electrons interacting with a radiation
field via the direct optical transition
mechanism. Under the action of a weak radiation field with
relatively low photon energies, the electronic response to the
radiation field is dominated by low-energy and small-momentum electrons.
Similar to a conventional semiconductor, the electronic transition accompanied by the absorption of photons
in a 3DDES is achieved mainly from occupied electronic states to empty states around the Fermi
energy without change of electron momentum. In this case, we can
limit ourselves to low-energy and small-momentum electrons and neglect
the high-order terms $B_\mathbf{K}\sim K^3\ll 1$ in Eq. (\ref{1}).
Thus, the block-diagonal form allows us to decouple the $4\times4$
matrix given by Eq. (\ref{1}) into a simplified $2\times2$ matrix, which reads
\begin{equation}\label{2}
H(\mathbf{K})=
\left(
\begin{array}{cc}
 \epsilon(\textbf{K})+M{_\mathbf{K}} & Ak_{+} \\
 Ak_{-} & \epsilon(\textbf{K})-M{_\mathbf{K}} \\
 \end{array}
\right).
\end{equation}
By solving the corresponding Schr\"{o}dinger equation, an
analytical expression for the eigenvalue and eigenfunction are
obtained, respectively, as
\begin{equation}\label{3}
E_\lambda(\mathbf{K})=\epsilon(\mathbf{K})+\lambda\sqrt{M_\mathbf{K}^2+A^{2}k^{2}},
\end{equation}
and
\begin{equation}\label{4}
\psi_{\lambda\mathbf{K}}(\mathbf{R})=|\mathbf{K},\lambda\rangle
=a_{\bf K}\left(
\begin{array}{cc}
 1 \\ b_{\bf K}
 \end{array}
\right)e^{i\mathbf{K}\cdot\mathbf{R}},
\end{equation}
with $\mathbf{R}=(x,y,z)$, $a_{\bf
K}=Ak(\xi_{\lambda\mathbf{K}}^2+A^{2}k^{2})^{-1/2}$, $b_{\bf
K}=\lambda\xi_{\lambda\mathbf{K}}/Ak_-$, $k=\sqrt{k_x^2+k_y^2}$,
$\xi_{\lambda\mathbf{K}}=\sqrt{M_\mathbf{K}+A^2k^2}-\lambda
M_\mathbf{K}$, and $\lambda=+1$ refers to conduction band and
$\lambda=-1$ to valance band. In this simplified
$\mathbf{k}\cdotp\mathbf{p}$ Hamiltonian, we can see that the two Dirac
points are still present at $k=0$ and $k_z=\pm k_c=\pm\sqrt{M_0/M_1}$
and the neglect of the $B_\mathbf{K}$ term in Eq. (\ref{1}) does not
significantly affect the electronic band structure
in the momentum and energy domains in which we are interested.
Therefore, from now on in this study we take the simplified
electron Hamiltonian given by Eq. (\ref{2}). Very recently,
we have applied this electronic band
structure for the investigation of the electronic transport properties
of Na$_3$Bi based 3DDES \cite{Yuan19} and good agreement between
theoretical and experimental \cite{Xiong15,Xiong16} results was achieved.

\subsection{\label{sec:leve22}{Carrier-photon interaction}}
In the present study, we use our previous theoretical approach that we
developed for investigating of the optoelectronic properties
for graphene \cite{Dong09} to calculate the optical conductivity for a 3DDES.
Here we consider electrons in a 3DDES interacting with a linearly
polarized radiation field as a continuous wave (CW). Because of the asymmetric
electronic energy spectrum, as given by Eq. (\ref{3}), the polarization
direction along the xy-plane (here taken along the x-direction) and
along the z-direction have to be considered separately. Including
the effect of the radiation field within the usual Coulomb gauge,
the carrier-photon interaction Hamiltonian can be obtained by taking
${\bf K}\to {\bf K}-e{\bf A}(t)$ in Eq. (\ref{2}) with ${\bf A}(t)$ being
the vector potential of the radiation field. Next, $H({\bf K}-e{\bf
A}(t))$ is expanded as $H({\bf K}-e{\bf A}(t))\simeq H({\bf K})+H'(t)$.
For the case of weak radiation intensity, we can neglect
the higher order effect with $A^2(t)$ terms. For the cases where light
polarization is along the $x$- and the $z$-direction of Na$_3$Bi, the
carrier-photon interaction Hamiltonian can be written respectively
as
\begin{equation}\label{5}
H_x'(t)=\frac{2eA(t)}{\hbar} \left(
\begin{array}{cc}
 (M_2-C_2)k_x & -A/2 \\
 -A/2 & -(M_2+C_2)k_x \\
 \end{array}\right), \\
\end{equation}
and
\begin{equation}\label{6}
H_z'(t)=\frac{2eA(t)k_z}{\hbar}
\left(
\begin{array}{cc}
 (M_1-C_1) & 0 \\
 0 & -(M_1+C_1) \\
 \end{array}\right),
\end{equation}
where $A(t)=(F_0/\omega){\rm sin}(\omega t)$ for a CW radiation field, $F_0$ is the electric
field strength of the light field, and $\omega$ is its frequency.
Because we limit ourselves to the case of weak
radiation field, we neglect the contribution from $F^2_0$ terms. As
we can see from Eqs. (\ref{5}) and (\ref{6}), when the radiation is polarized linearly
along different directions the carrier-photon interaction
Hamiltonian takes different functional forms.

The first-order steady-state electronic transition rate induced by
carrier-photon interactions via photon absorption scattering in Eq. (\ref{5}) and
Eq. (\ref{6}) can be obtained using the Fermi's golden rule, which reads
\begin{equation}\label{7}
\begin{split}
W^j_{\lambda\lambda'}(\mathbf{K},\mathbf{K}')=&\frac{2\pi}{\hbar}\bigg
(\frac{eF_0}{\omega\hbar}\bigg)^2|U^j_{\lambda\lambda'}(\mathbf{K})|^2
\delta_{\mathbf{K',K}}\\
&\times\delta[E_{\lambda'}(\mathbf{K}')-E_\lambda(\mathbf{K})-\hbar\omega]
\end{split}
\end{equation}
with $j=x$ or $z$. This is the probability for scattering of a carrier from a
state $|\mathbf{K},\lambda\rangle$ to a state
$|\mathbf{K'},\lambda'\rangle$ induced by the interaction with the
linearly polarized radiation field, accompanied by the absorption of
a photon with energy $\hbar\omega$. Here,
\begin{equation}\label{8}
|U^j_{\lambda\lambda'}(\mathbf{K})|^2={|G^j_{\lambda\lambda'}({\bf
K})|^2 \over [\xi_{\lambda\mathbf{K}}^2+A^{2}k^{2}]
[\xi_{\lambda'\mathbf{K}}^2+A^{2}k^{2}]},
\end{equation}
with
\begin{equation}\nonumber
\begin{split}
G^x_{\lambda\lambda'}({\bf K})&=A^2(\lambda
k_+\xi_{\lambda\mathbf{K}}+\lambda'k_-\xi_{\lambda'\mathbf{K}})/2 \\
&-k_x[A^2k^2(M_2-C_2)-\lambda\lambda'(M_2+C_2)\xi_{\lambda\mathbf{K}}\xi_{\lambda'\mathbf{K}}],
\end{split}
\end{equation}
and
\begin{equation}\nonumber
G^z_{\lambda\lambda'}({\bf
K})=k_z[A^2k^2(M_1-C_1)-(M_1+C_1)\lambda\lambda'\xi_{\lambda\mathbf{K}}\xi_{\lambda'\mathbf{K}}].
\end{equation}
\subsection{\label{sec:leve23}{Optical conductivity}}
In this work, we employ the Boltzmann equation as the governing
transport equation to study the response of carriers in Na$_3$Bi
to an applied light field. The semi-classical Boltzmann equation
can be written as
\begin{equation}\label{9}
\frac{\partial f_\lambda(\mathbf{K})}{\partial
t}=g_s\sum_{\lambda',\mathbf{K}'}
[F^j_{\lambda\lambda'}(\mathbf{K},\mathbf{K'})-F^j_{\lambda'\lambda}(\mathbf{K'},\mathbf{K})]
\end{equation}
where $g_s=2$ counts for for spin degeneracy, $f_\lambda(\mathbf{K})$ is the
momentum-distribution function for a carrier at state $|\mathbf{K},\lambda\rangle$ and
$F^j_{\lambda\lambda'}(\mathbf{K},\mathbf{K'})=f_\lambda(\mathbf{K})[1-f_{\lambda'}(\mathbf{K'})]
W^j_{\lambda\lambda'}(\mathbf{K},\mathbf{K}')$. Because the
radiation field has already been included within the electronic
transition rate, the force term induced by the light field does not
appear in the drift term on the left-hand side of the Boltzmann
equation. Eq. (\ref{9}) with the electronic
transition rate $W^j_{\lambda\lambda'}(\mathbf{K},\mathbf{K}')$
given by Eq. (\ref{7}) cannot be solved analytically. In this study we
employ the usual balance-equation approach \cite{Xu91,Xu05} to
approximately solve the problem. By multiplying
$g_s\sum_{\lambda,\mathbf{K}}E_\lambda({\bf K})$ to both sides of
the Boltzmann equation, Eq. (\ref{9}), we obtain the energy-balance
equation. From the energy-balance equation, we get the energy
transfer rate for a carrier in 3DDES:
$P^j(\omega)=g_s\sum_{\lambda,{\bf K}}E_\lambda({\bf K})\partial
f_\lambda({\bf K})/\partial
t=\sum_{\lambda\lambda'}P^j_{\lambda\lambda}(\omega)$ with
\begin{equation}\label{10}
P^j_{\lambda\lambda'}(\omega)=4\hbar\omega\sum_{\mathbf{K},\mathbf{K}'}
F^j_{\lambda\lambda'} (\mathbf{K},\mathbf{K'}).
\end{equation}
Knowing the electronic energy
transfer rate, we can calculate the optical conductivity via
\cite{Ec06}: $\sigma_{jj}(\omega)=P(\omega)/(2F_0^2)$. Once the
optical conductivity is obtained, we can calculate other optical
coefficients such as the absorption and the transmission coefficients by
using basic laws of electrodynamics. The optical conductivity
obtained from the energy-balance equation is given by
\begin{equation}\label{11}
\sigma_{jj}(\omega)=\sum_{\lambda,\lambda'}\sigma_{jj}^{\lambda\lambda'}
(\omega)=\frac{2\hbar\omega}{F_0^2}\sum_{\lambda,\lambda'}
\sum_{\mathbf{K},\mathbf{K}'}F^j_{\lambda\lambda'}(\mathbf{K},\mathbf{K'}).
\end{equation}
Because $F^j_{\lambda\lambda'}(\mathbf{K},\mathbf{K'})\sim F_0^2$
(see Eq. (\ref{7})), $\sigma_{jj}(\omega)$ is independent of the radiation
intensity $F_0$ in case $F_0$ is sufficiently weak.

One of the major advantages of the balance-equation approach is that
we can circumvent the difficulties of solving the Boltzmann equation
directly by using a specific form of the distribution function. In
this study, we assume that the momentum distribution function for
carriers in Na$_3$Bi can be described by a statistical energy
distribution such as the Fermi-Dirac function, namely we take
$f_\lambda ({\bf K})\simeq f_\lambda[E_\lambda({\bf K})]$ with
$f_+(x)=[1+e^{(x-E_F^e)/k_BT}]^{-1}$ or
$f_-(x)=[1+e^{(x-E_F^h)/k_BT}]^{-1}$ being respectively the
Fermi-Dirac function for electrons or holes, where $E_F^e$ and
$E_F^h$ are chemical potentials (or Fermi energies at $T\to 0$) for
electrons and holes, respectively. It should be noted that similar to a
semiconductor, for a n-type Na$_3$Bi subjected to a radiation field,
photon-induced carriers are possible. As a result, the
chemical potentials for electrons and holes can be different.

After considering the effect of the broadening of the scattering
states due to energy relaxation, i.e., by taking $\delta(E)\to
(E_\tau/ \pi)(E^2+E_\tau^2)^{-1}$ (where $E_\tau= \hbar/\tau$) in
Eq. (\ref{7}) with $\tau$ being the energy relaxation time, we can calculate
the optical conductivity induced by different transition
channels. For the situation where the radiation field is polarized
along the $x$-direction, the optical conductivity induced
by intra-band transitions is obtained as
\begin{equation}\label{12}
\begin{split}
\sigma_{xx}^{\lambda\lambda}(\omega)=&\frac{\sigma_0}{A_\omega^2}\frac{\omega\tau}
{1+(\omega\tau)^2}\int_{0}^{\infty}dk_z\int_{0}^{\infty}dk
k^3{\cal G}^x_{\lambda\lambda}(\mathbf{K})\\
&\times
f_{\lambda}(E_\lambda(\mathbf{K}))[1-f_{\lambda}(E_{\lambda}(\mathbf{K}))],
\end{split}
\end{equation}
where $\lambda=\lambda'=\pm1$ refer to conduction and valance band,
$\sigma_0=e^2/\hbar$, $A_\omega=\pi\hbar\omega$, and
\begin{equation}\nonumber
{\cal G}^x_{\lambda\lambda}(\mathbf{K})=
\big[{A^2[k^2(M_2-C_2)-\lambda\xi_{\lambda\mathbf{K}}]-(M_2+C_2)\xi_{\lambda\mathbf{K}}^2\over
\xi_{\lambda\mathbf{K}}^2+A^{2}k^{2}}\big]^2.
\end{equation}
For inter-band transitions, we have $\sigma_{xx}^{+-}(\omega)\to 0$
because of the energy conservation required for electronic transitions and
\begin{equation}\label{13}
\begin{split}
\sigma_{xx}^{-+}(\omega)=&\frac{\sigma_0 A^2
\omega\tau}{A_\omega^2}\int_{0}^{\infty}dk_z\int_{0}^{\infty}dk k
{\cal
G}^x_{-+}(\mathbf{K})\\
&\times\frac{f_{-}(E_-(\mathbf{K}))[1-f_{+}(E_{+}(\mathbf{K}))]}{1+
(\omega-2\sqrt{M_{\mathbf{K}}^2+A^2k^2}/\hbar)^2\tau^2},
\end{split}
\end{equation}
with
\begin{equation}\nonumber
{\cal G}^x_{-+}(\mathbf{K})=1+\frac{(2M_2k^2+M_{\mathbf{K}})^2}
{M_{\mathbf{K}}^2+A^2k^2}.
\end{equation}
For the situation of light polarization along the $z$-direction,
the optical conductivity induced by the intra-band transitions is obtained
as
\begin{equation}\label{14}
\begin{split}
\sigma_{zz}^{\lambda\lambda}(\omega)=&\frac{2\sigma_0}{A_\omega^2}\frac{\omega\tau}{1+(\omega\tau)^2}
\int_{0}^{\infty} \int_{0}^{\infty}dkdk_z k k_z^2\\
&\times {\cal G}^z_{\lambda\lambda}(\mathbf{K})
f_{\lambda}(E_\lambda(\mathbf{K}))[1-f_{\lambda}(E_{\lambda}(\mathbf{K}))],
\end{split}
\end{equation}
with
\begin{equation}\nonumber
{\cal
G}^z_{\lambda\lambda}(\mathbf{K})=\bigl[\frac{A^2k^2(M_1-C_1)-(M_1+C_1)\xi_{\lambda\mathbf{K}}^2}
{\xi_{\lambda\mathbf{K}}^2+A^{2}k^{2}}\bigr]^2.
\end{equation}
For inter-band transitions, we have $\sigma_{zz}^{+-}(\omega)\to 0$
and
\begin{equation}\label{15}
\begin{split}
\sigma_{zz}^{-+}(\omega)=&{2\sigma_0\over
A_\omega^2}{A^2M_1^2\omega\tau}
\int_{0}^{\infty}\int_{0}^{\infty}{dk dk_z k^3k_z^2\over M^2_{\bf K}+A^2k^2}\\
&\times {f_{-}(E_-(\mathbf{K}))[1-f_{+}(E_{+}(\mathbf{K}))]\over 1+
(\omega-2\sqrt{M_{\mathbf{K}}^2+A^2k^2}/\hbar)^2\tau^2}.
\end{split}
\end{equation}
Using these equations, we can evaluate the contributions from
different transition channels to the optical conductivity or
the optical absorption.

\section{\label{sec:leve3}{RESULTS AND DISCUSSION}}
\begin{figure*}[t]
  \centering
  \subfigure{\label{fig1a}
  \begin{minipage}{6cm}
  \centering
    \includegraphics[width=6.5cm]{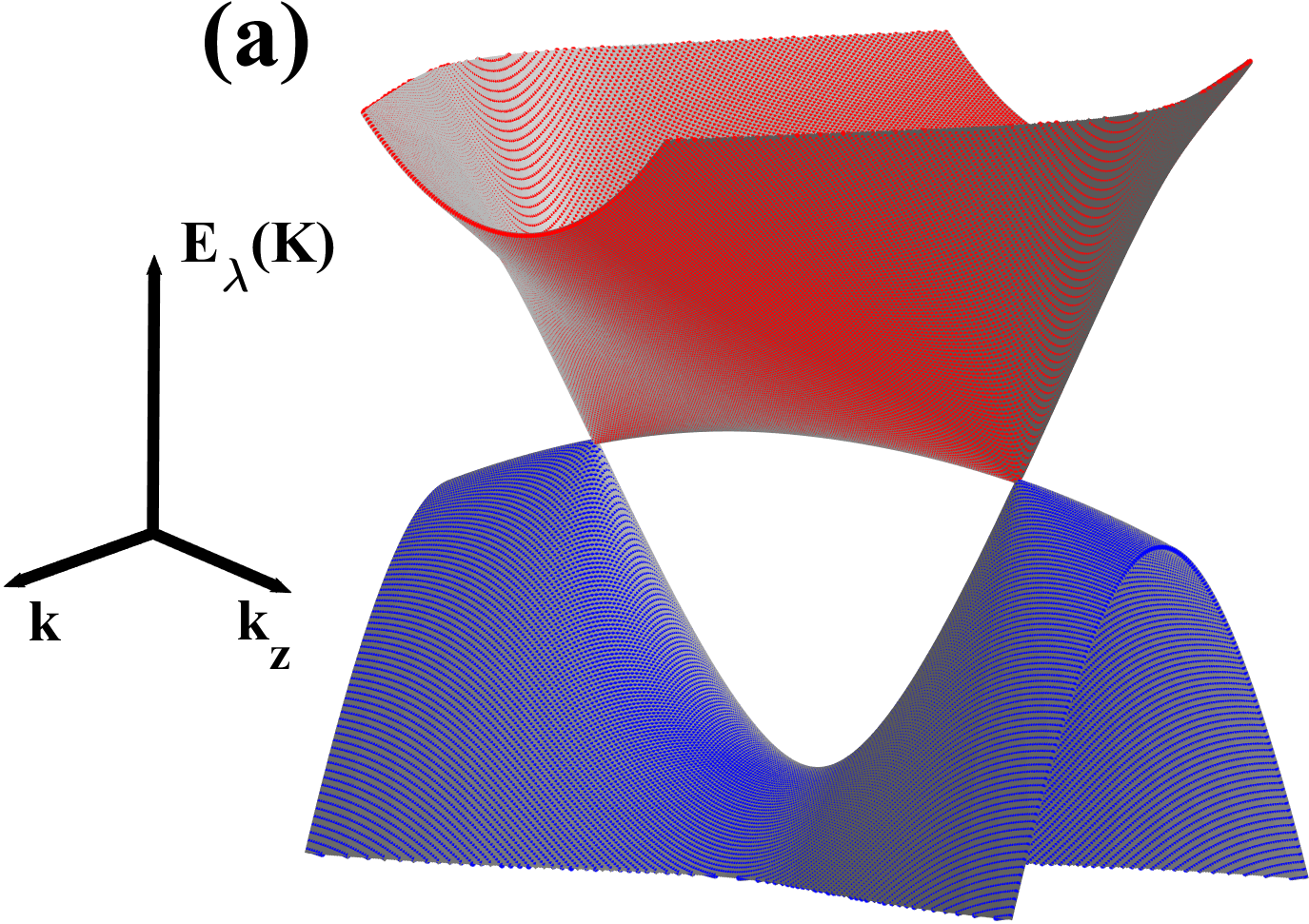}
  \end{minipage}}
  \subfigure{\label{fig1b}
  \begin{minipage}{11.5cm}
  \centering
    \includegraphics[width=12cm]{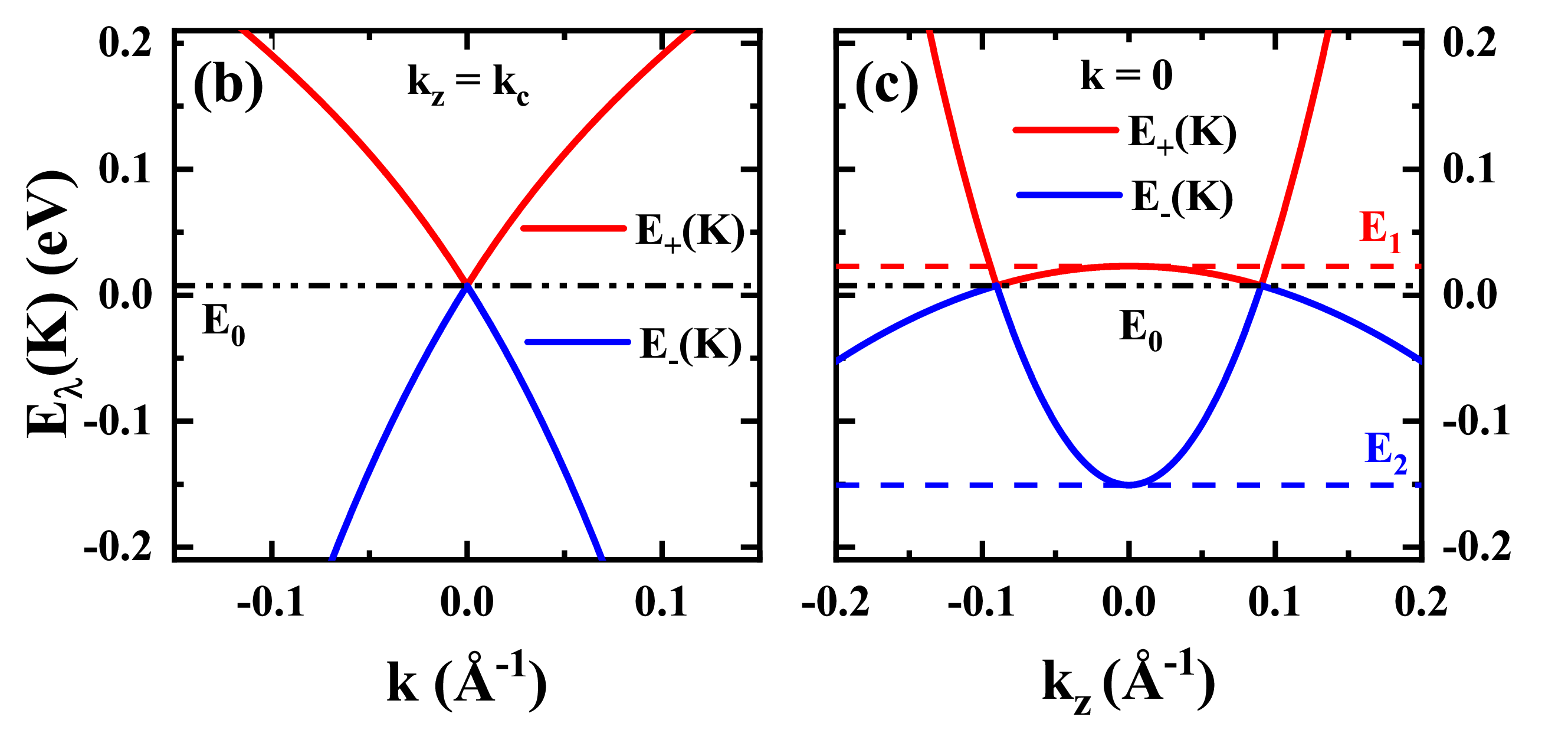}
  \end{minipage}}
  \caption{(a) The conduction (red) and valance (blue) band energies in a Na$_3$Bi-based 3DDES as a
  function of $k$ and $k_z$, as given by Eq. (\ref{3}).
  (b) The electron energy as a function of $k$ at $k_z=k_c$. The band crossing point is at $E_0\simeq7.6$ meV
  for $k=0$ and $k_z=\pm k_c$ (dashed-dotted line). (c) The electron energy as a function
  of $k_z$ at $k=0$. The top of the Berry curvature in the conduction band
  is $E_1\simeq23$ meV and the bottom of the Berry curvature in the valance band is $E_2\simeq-151$
  meV, which are both located at $k=0$ and $k_z=0$.}\label{fig1}
\end{figure*}
\begin{figure*}[t]
  \includegraphics[width=14cm]{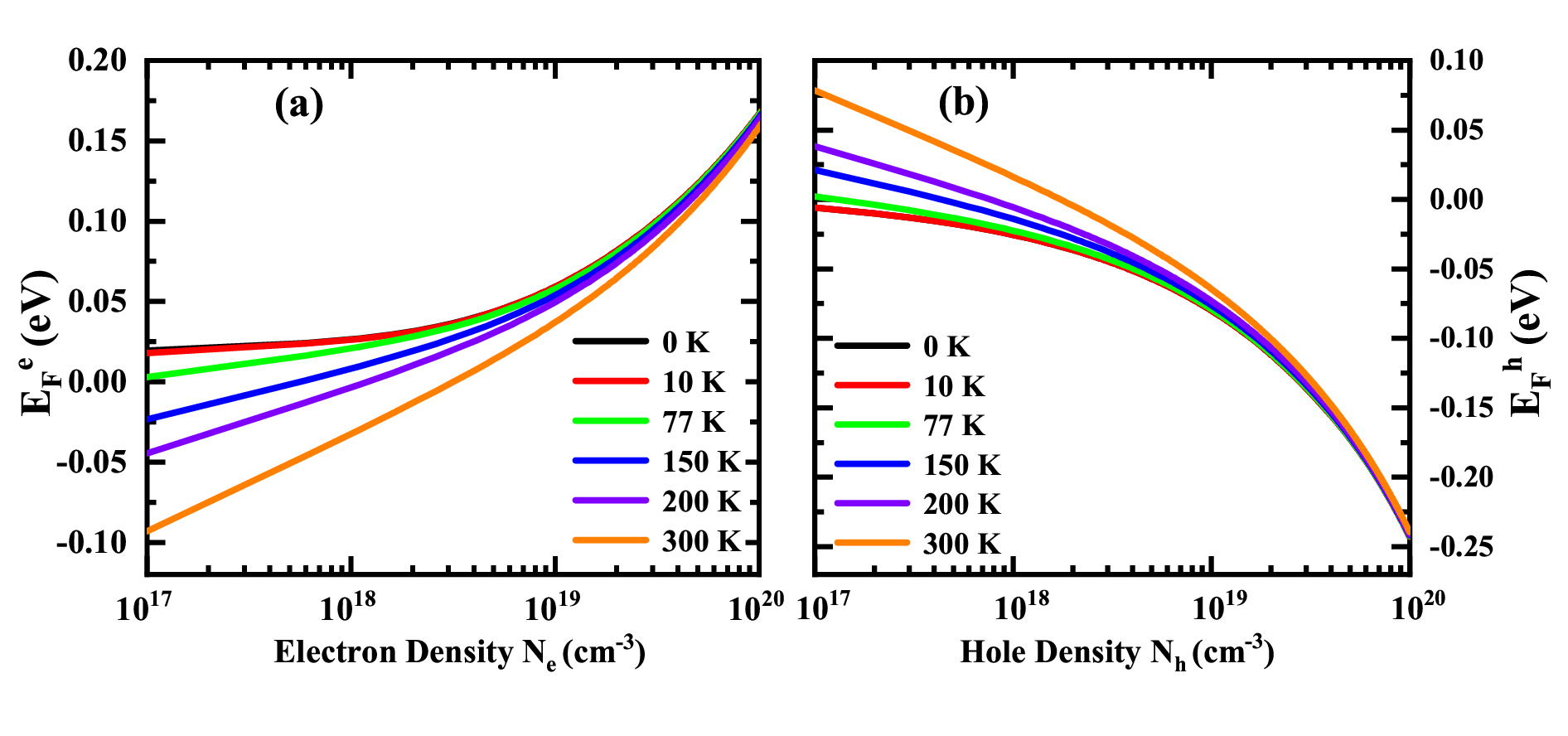}
  \caption{The chemical potential for electrons $E_F^e$ (a) and for holes (b) in a Na$_3$Bi based 3DDES as
  a function of electron density $N_e$ and hole density $N_h$ at different temperatures as indicated.
  The results for $T=0$ and $T=10$ K almost coincide. }\label{fig2}
\end{figure*}
\begin{figure}[t]
 \includegraphics[width=8cm]{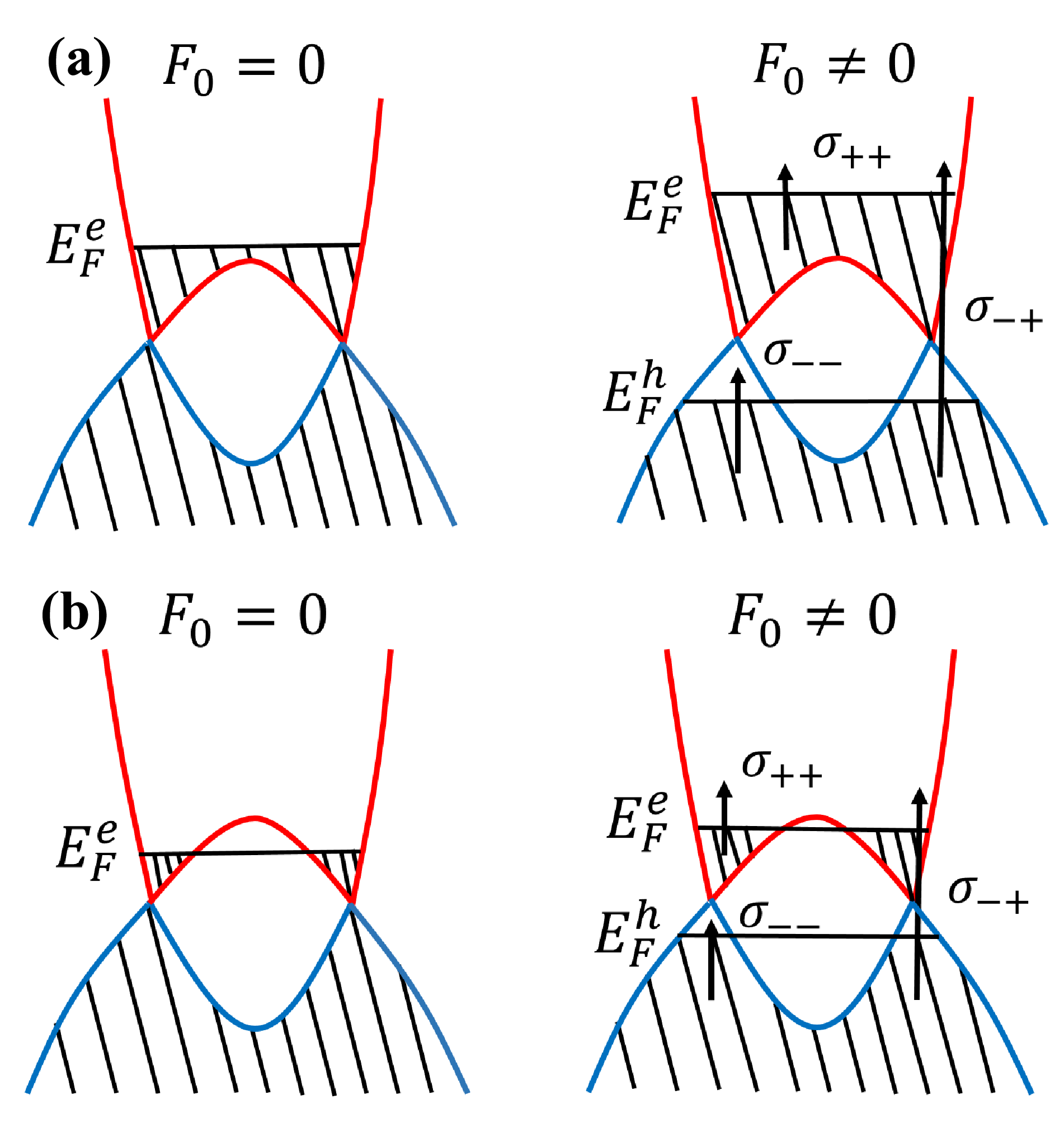}
 \caption{Schematic diagram of the electronic transitions accompanied by the absorption of a photon
  in a n-type 3DDES. Here, $F_0=0$ and $F_0\neq 0$ refer respectively to the cases in the absence
  and presence of the radiation field. $E_F^e$ and $E_F^h$ are respectively the Fermi energy for
  electrons and holes, and the up-arrows stand for different transition channels with corresponding
  optical conductivities $\sigma_{\lambda\lambda'}$ induced by intra- and inter-band transition events.
  Two situations for Fermi energy $E_F^e$ outside [in (a)] and inside [in (b)] the Berry curvature
  regime are indicated. $E_F^h$ is always inside the Berry curvature regime in the valance band.}\label{fig3}
\end{figure}
For numerical calculations, we take the band
parameters for Na$_3$Bi as \cite{Yuan19}: $C_0=-0.06382$ eV, $C_1=8.7536$ eV$\AA^2$, $C_2=-8.4008$ eV$\AA^2$,
$M_0=-0.08686$ eV, $M_1=-10.6424$ eV$\AA^2$, $M_2=-10.3610$ eV$\AA^2$, and $A=2.4598$ eV$\AA$.
These parameters were determined by fitting the energy spectrum of the effective
Hamiltonian given by Eq. (\ref{1}) with those obtained from $ab$ $initio$ calculations
\cite{Wang12}. Furthermore, it has been shown
experimentally that in a Na$_3$Bi 3DDES, the electronic relaxation time
is different for samples with different carrier densities \cite{Xiong16}. Thus, we take the following
$\tau=1.0 $ ps, $1.49$ ps, $1.94$ ps, $2.55$ ps and $6.71$ ps obtained experimentally \cite{Xiong16} in our calculations.

By definition, the condition of carrier number conservation is given by:
$N_\lambda=g_s\sum_\mathbf{K}f_\lambda[E_\lambda(\mathbf{K})]$ with
$N_\lambda$ being the electron or hole density. Using this
condition, we can determine the corresponding chemical potential
$E_F$ for electrons and holes at finite-temperature, which reads
\begin{equation}\label{16}
N_e=\frac{1}{\pi^2}\int_0^\infty dk_z\int_0^\infty
dkkf_+[E_+(\mathbf{K})],
\end{equation}
for conduction band and
\begin{equation}\label{17}
N_h=\frac{1}{\pi^2}\int_0^\infty dk_z\int_0^\infty
dkk[1-f_-[E_-(\mathbf{K})]],
\end{equation}
for valance band. Thus, the chemical potential $E_F^\lambda$ for electrons
and holes can be obtained numerically from respectively Eqs. (\ref{16}) and
(\ref{17}) for given carrier densities $N_e$ and $N_h$ at finite-temperature $T$.
It should be noted that for n-type Na$_3$Bi, if $N_0$ is the electron density
in the absence of the radiation field (or dark density), the electron density in
the presence of the radiation is $N_e=N_0+\Delta N_e$, where $\Delta
N_e$ is the density of photo-excited
electrons. Under the condition of charge number conservation,
$\Delta N_e=N_h$ is the hole density in the presence of light
irradiation.

The features of the conduction band structure of
Na$_3$Bi 3DDES have been discussed previously \cite{Yuan19}
and will therefore not be repeated. Because inter-band electronic transitions
are involved in the present study, here we present and discuss the conduction
and valence band structure for semi-metal Na$_3$Bi. In Fig. \ref{fig1},
we show the energy dispersion relation of a Na$_3$Bi-based 3DDES, given by
Eq. (\ref{3}) for conduction and valance bands. In Fig. \ref{fig1}(a), the 3D
plot is shown for $E_\lambda ({\bf K}) $ as function of $k$ and $k_z$. We see
that the electron energy has a different dependence on $k$ and $k_z$ for both
conduction and valance bands, showing the asymmetric nature of the electron
energy spectrum in the $xy$-plane and along the $z$-direction in Na$_3$Bi 3DDES.
In order to see more clearly the Dirac points and the Berry curvature in the
energy bands, we show the electron energy as a function of $k$ at a fixed
$k_z=k_c$ in Fig. \ref{fig1}(b) and as a function of $k_z$ at $k=0$ in
Fig. \ref{fig1}(c). We note that in the theoretical model, the bottom of
the conduction band or the top of the valance band is at $E_0=7.6$ meV, which
is reached by taking $k=0$ and $k_z=\pm k_c$ (see Fig. \ref{fig1}(b)).
Thus, the energy reference in this study is from $E_0=7.6$ meV. From
Fig. \ref{fig1}, we notice the following special features: i) two Dirac points can be found at the bottom of the
conduction band and the top of the valance band when $k_z=\pm k_c$ and $k=0$.
ii) Around the Dirac points, the electron energies for conduction and valance
bands are asymmetric and depend nonlinearly on $k$ and $k_z$. These features
are in sharp contrast to those in graphene. iii) The electronic energy spectrum
show arch-bridge-like Berry curvatures in both conduction and valance bands. The
effect of the Berry curvature is much stronger in the valance band. The top of the
Berry curvature is at $E_1\simeq23$ meV in the conduction band and the bottom
of the Berry curvature is at $E_2\simeq-151$ meV in the valance band, which are
both reached at $k=k_z=0$ (see Fig. \ref{fig1}(c)). These features for a 3DDES
differ significantly from those for graphene and from conventional 3D electron gas systems.

In Fig. \ref{fig2}, we show the chemical potential $E_F^\lambda$ as a function
of carrier density at different temperatures, where $N_e$ and $N_h$ are respectively
the electron and hole density. If there is no conducting carriers in the system,
$E_F^e$ and $E_F^h$ reach the Dirac points at about $E_0=7.6$ meV at $T\to 0$.
Similar to a conventional semiconductor, the Fermi energy for electrons (holes)
in a 3DDES increases (decreases) with increasing carrier density but decreases
(increases) with increasing temperature. The results shown in Fig. \ref{fig1}
and Fig. \ref{fig2} suggest that when $E_F^e$ is smaller than $E_0+E_1=30.6$ meV,
the electrons are mainly located in the conduction band of the
energy range with Berry curvature, which corresponds to an electron density
$N_e\sim4\times10^{18}$ cm$^{-3}$ \cite{Yuan19}. For n-type Na$_3$Bi, because of
the low hole density generated mainly via optical excitation and of the strong
Berry curvature effect, the holes are always located in the valance band energy
regime with  Berry curvature. Experimentally, the electron density in a n-type
Na$_3$Bi sample can be altered through post-annealing \cite{Xiong15,Xiong16}.
Thus, one can tune the Fermi level inside and outside the Berry curvature regime
in order to see anomalous features in, e.g., the conductivity \cite{Xiong15,Xiong16}.

To gain an in-depth understanding of the mechanism for optical response of
carriers in a 3DDES, in Fig. \ref{fig3} we present schematic diagrams of the
electronic transition channels in the absence and presence of a radiation field.
In the absence of the radiation field (i.e., $F_0=0$), the valance band is fully
occupied in a n-type 3DDES and the Fermi energy for electrons crosses the conduction
band. In the presence of the radiation field (i.e., $F_0\neq 0$), electrons in the
valance band are excited optically from the valance band to the conduction band.
Thus, photon-generated electrons are added to the conduction band and the holes
are generated in the valance band along with the creation of the Fermi energy for
holes in the valance band. Like in other electronic systems, the optical response
of carriers in a 3DDES can be achieved through electronic transitions from the lower
occupied states to the higher empty states due to the absorption of a photon. We
see from Fig. \ref{fig3} that the main difference between the different cases where
the Fermi energy $E_F^e$ is outside [in (a)] or inside [in (b)] the Berry curvature
regime in the conduction band is the density-of-states for initial and final states
of the electronic transitions. In the presence of a light field the intra-band
electronic transition via optical absorption can be achieved not only in the
conduction band via the channel $\sigma_{++}$ but also in the valence band via the
channel $\sigma_{--}$. The intra-band transitions are a consequence of direct
optical absorption in the 3DDES. Because Na$_3$Bi is a gapless
3DDES, the electrons in the valence band can be more easily excited optically into
the conduction band in contrast to a conventional semiconductor. Thus, there is a
strong inter-band transition channel $\sigma_{-+}$ (in Fig. \ref{fig3}) in Na$_3$Bi-based 3DDES.
These features are similar to those in the 2D Dirac system such as graphene \cite{Dong09}.
Since optical absorption is achieved by electronic
transitions from occupied states to empty states, together with the presence of
the Moss-Burstein effect \cite{Bu54} or the Pauli blockade effect \cite{Kr06},
intra-band transitions require less photon energy $\hbar\omega$,
whereas a relatively larger photon energy is needed for inter-band transitions.
\begin{figure}[t]
\includegraphics[width=8cm]{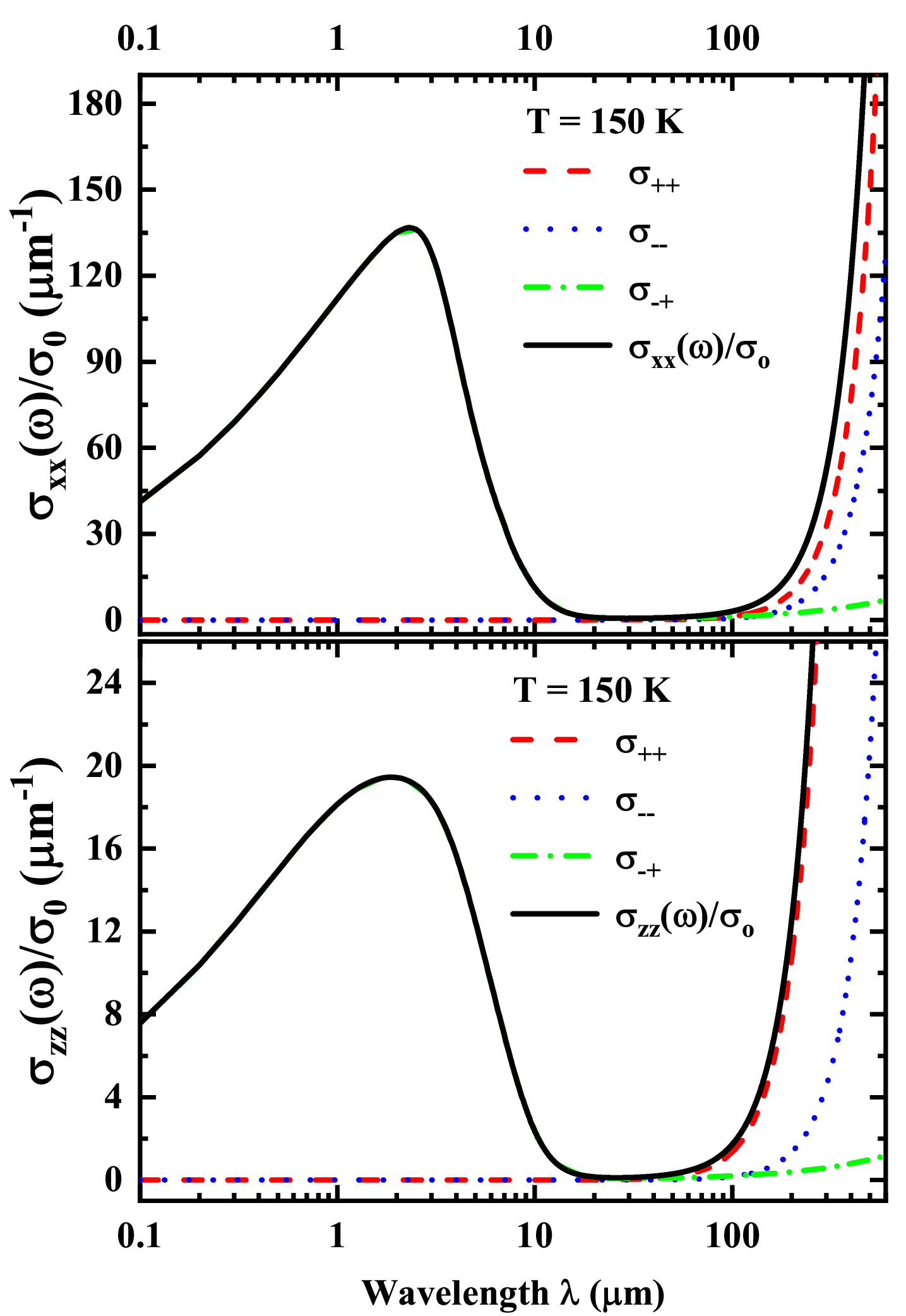}
\caption{Contributions from different electronic transition channels to optical conductivity
for light radiation polarized linearly along the $x$-direction $\sigma_{xx}(\omega)$ (upper panel) and along the $z$-direction $\sigma_{zz}(\omega)$
(lower panel) at the fixed temperature $T=150$ K, electronic relaxation time $\tau=6.71$ ps, electron density $N_e=2.0\times10^{19}$ cm$^{-3}$, and hole density $N_h=2.0\times10^{18}$ cm$^{-3}$. Here $\sigma_0=e^2/\hbar$.}\label{fig4}
\end{figure}

In Fig. \ref{fig4}, we show the contributions from different electronic transition
channels to the optical conductivity induced by light irradiation with different
polarization directions in a Na$_3$Bi based 3DDES, where $\sigma_{xx}(\omega)$
and $\sigma_{zz}(\omega)$ are shown respectively in the upper and lower panels at
fixed temperature $T=150$ K, electronic relaxation time $\tau=6.71$ ps, electron
density $N_e=2.0\times10^{19}$ cm$^{-3}$, and hole density $N_h=2.0\times10^{18}$ cm$^{-3}$.
From Fig. \ref{fig4}, we notice: i) the inter-band transitions contribute to
the optical conductivity in the short-wavelength regime and the intra-band transitions give
rise to the long-wavelength optical conductivity in both $\sigma_{xx}(\omega)$ and
$\sigma_{zz}(\omega)$. ii) The short-wavelength optical conductivity in a 3DDES increases
with radiation wavelength when $\lambda<$ 2 $\mu m$. An optical absorption peak can be
observed. This phenomenon differs from graphene in which the optical conductivity is
universal \cite{Ku08,Wang08}, namely $\sigma(\omega)=e^2/(4\hbar)$ and does not depend
on the radiation frequency, temperature and sample parameters such as electron density
and scattering strength. iii) In the short-wavelength regime, $\sigma_{xx}(\omega)$
is about 5 times larger than $\sigma_{zz}(\omega)$. This suggests that stronger optical
absorption can be achieved when the radiation field is polarized linearly along the
2D-plane of a 3DDES. The optical absorption peak for $\sigma_{xx}(\omega)$ is sharper
than that for $\sigma_{zz}(\omega)$. From the electron energy spectrum shown in
Fig. \ref{fig1}, we see that $E_\lambda ({\bf K})$ for a Na$_3$Bi based 3DDES depends
differently on $k$ and $k_z$. Because the Berry curvature appears at $k_z=\pm k_c$,
the density-of-states (DoS) for electrons along the $z$-direction is smaller than
in the $xy$ plane, so that $\sigma_{xx}$ is larger than $\sigma_{zz}$. iv) More
interestingly, both $\sigma_{xx}(\omega)$ and $\sigma_{zz}(\omega)$ show an optical
absorption window in the intermediate wavelength regime (about $5$ $\mu m<\lambda<400$ $\mu m$
for $\sigma_{xx}$ and $5$ $\mu m<\lambda<200$ $\mu m$ for $\sigma_{zz}$), because
the optical absorption coefficient is proportional to optical conductivity. This
effect is induced by the blocking of the optical absorption channels due to inter-
and intra-band electronic transitions as shown in Fig. \ref{fig3}. Thus, similar to
the optical absorption window observed experimentally for graphene \cite{Ku08,Fa07,Ry09},
this window in a 3DDES is also caused by the different energy requirements for the
inter- and intra-band electronic transition channels. v) In a Na$_3$Bi 3DDES,
the high-frequency edge of the optical absorption window is located between
$5-10$ $\mu m$, different from that in graphene in which the high-frequency edge
of the optical absorption window is at $0.1$ $\mu m$ \cite{Dong09}. vi) In the
long-wavelength regime ($\lambda> 100$ $\mu m$), the intra-band electronic transitions
contributes mainly to the optical conductivity where both $\sigma_{xx}(\omega)$ and
$\sigma_{zz}(\omega)$ increases sharply with the radiation wavelength. This is similar
to the dependence of the optical conductivity on the radiation wavelength in the Durde
optical conductivity of free electrons.

\begin{figure}[t]
\includegraphics[width=8cm]{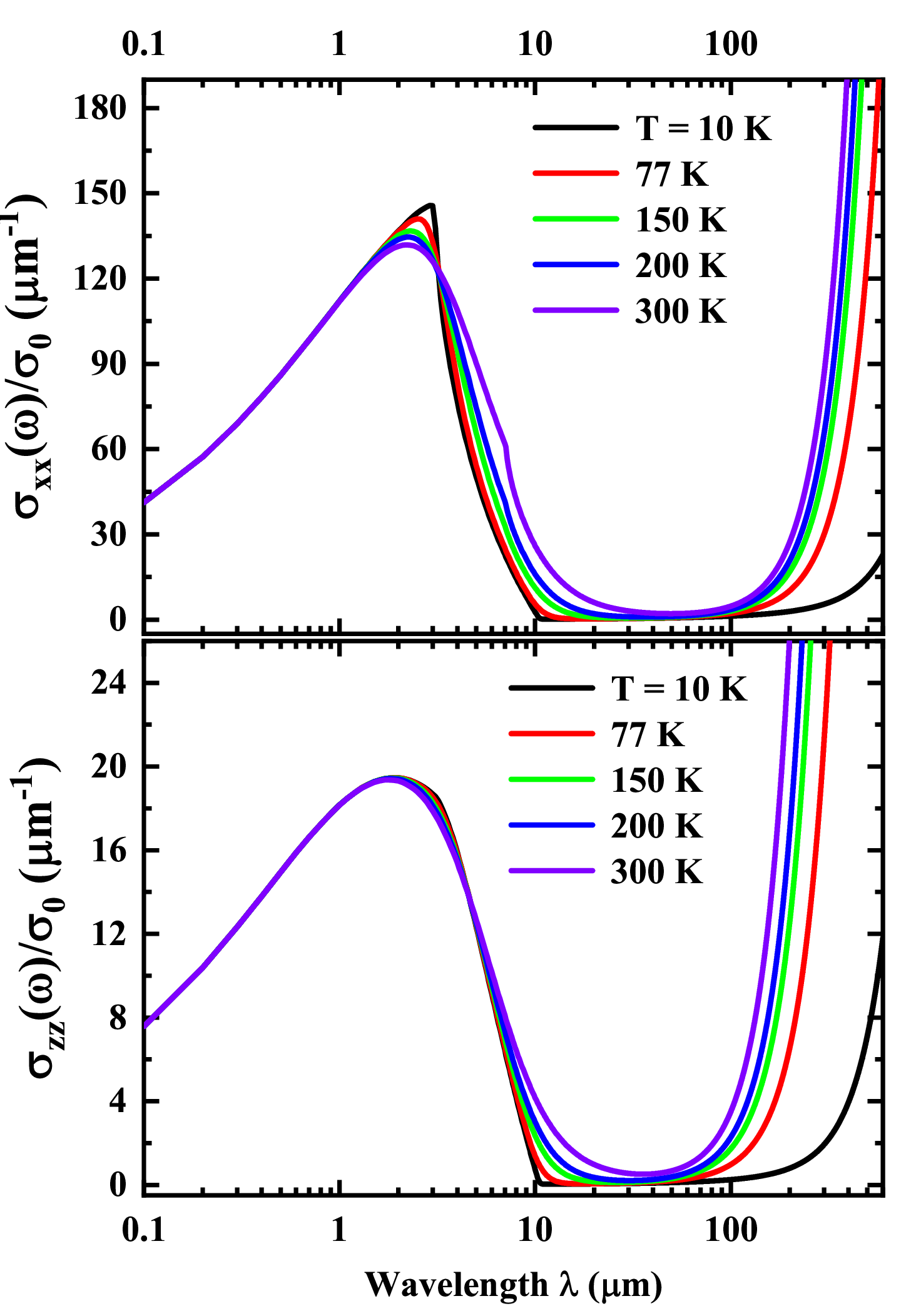}
\caption{Optical conductivities $\sigma_{xx}(\omega)$ (upper panel) and $\sigma_{zz}$ (lower panel) as a function
of radiation wavelength at the fixed electronic relaxation time $\tau=6.71$ ps and
carrier densities $N_e=2.0\times10^{19}$ cm$^{-3}$ and $N_h=2.0\times10^{18}$ cm$^{-3}$
for different temperatures $T=10$ K (black curve), $77$ K (red curve),
$150$ K (green curve), $200$ K (blue curve), and $300$ K (purple curve).}\label{fig5}
\end{figure}
In Fig. \ref{fig5}, we show the optical conductivities $\sigma_{xx}(\omega)$ and $\sigma_{zz}(\omega)$
as a function of radiation wavelength $\lambda$ for fixed carrier densities and electronic
relaxation time at different temperatures. We find that in the short-wavelength regime, both
$\sigma_{xx}(\omega)$ and $\sigma_{zz}(\omega)$ do not depend on temperature. The peaks of the
optical absorption in $\sigma_{xx}(\omega)$ and $\sigma_{zz}(\omega)$ depends weakly on
temperature, but the width of the optical absorption window becomes narrow with increasing
temperature. This is because $E_F^e$ decreases and $E_F^h$ increases with increasing
temperature (see Fig. \ref{fig2}). In such a case, the energy gap $E_F^e-E_F^h$ for
inter-band transitions decreases with increasing temperature due to Pauli blockade
effect \cite{Bu54,Kr06}, so that the photon energy required for inter-band transitions
decreases with increasing temperature (see Fig. \ref{fig3}). Furthermore, the decrease
in $E_F^e$ and the increases in $E_F^h$ with increasing temperature imply that intra-band
electronic transitions need a slightly larger photon energy to overcome the effect of
thermal broadening of the electron distribution function. It is interesting to notice
that a little kink of $\sigma_{xx}(\omega)$ can be seen at $T=300$ K at the high frequency
edge of the optical absorption window. When $N_e=2.0\times10^{19}$ cm$^{-3}$, $E_F^e$
for $T=0$ K is well above the Berry curvature regime in conduction band (see Fig. \ref{fig2}).
With increasing $T$, $E_F^e$ decreases and $E_F^e$ is located inside the Berry curvature
regime. The change of the electron DoS due to the effect of the Berry curvature is the
reason for the observation of this kink in $\sigma_{xx}(\omega)$ at $T=300$ K. Moreover,
it can be seen that there is a crossover of both $\sigma_{xx}(\omega)$ and $\sigma_{zz}(\omega)$
at the high frequency edge of the optical absorption window. Similar effect was observed theoretically
in graphene-based 2D Dirac systems \cite{Dong09}.
\begin{figure}[t]
  \includegraphics[width=8cm]{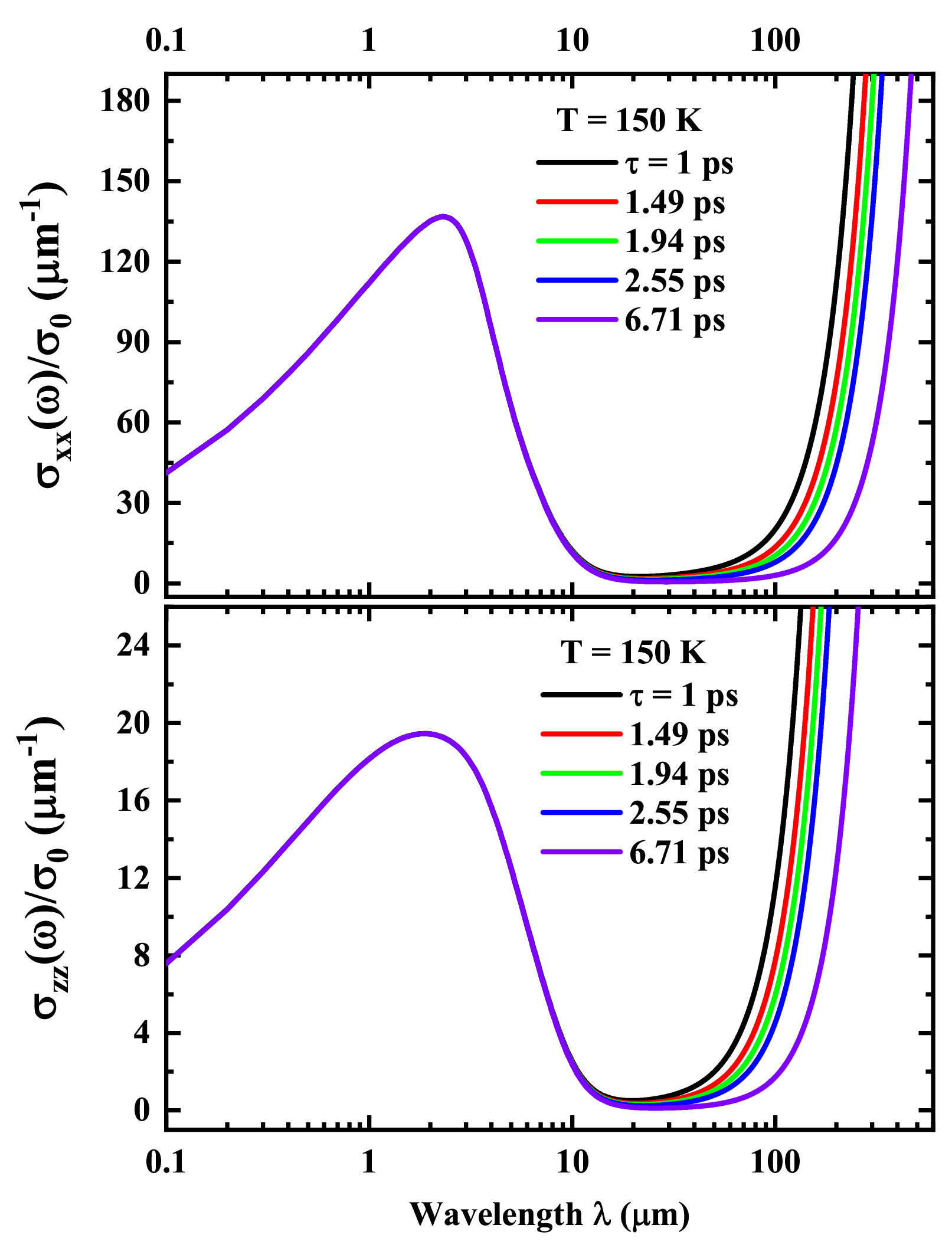}
  \caption{The dependence of $\sigma_{xx}(\omega)$ (upper panel) and $\sigma_{zz}(\omega)$ (lower panel) on radiation wavelength
  at a fixed temperature $T=150$ K and fixed carrier densities
  $N_e=2.0\times10^{19}$ cm$^{-3}$ and $N_h=2.0\times10^{19}$ cm$^{-3}$ for
  different electronic relaxation times $\tau=1$ ps (black curves), $1.49$ ps (red curves),
  $1.94$ ps (green curves), $2.55$ ps (blue curves), and
  $6.71$ ps (purple curves).}\label{fig6}
\end{figure}
\begin{figure}[t]
  \includegraphics[width=8cm]{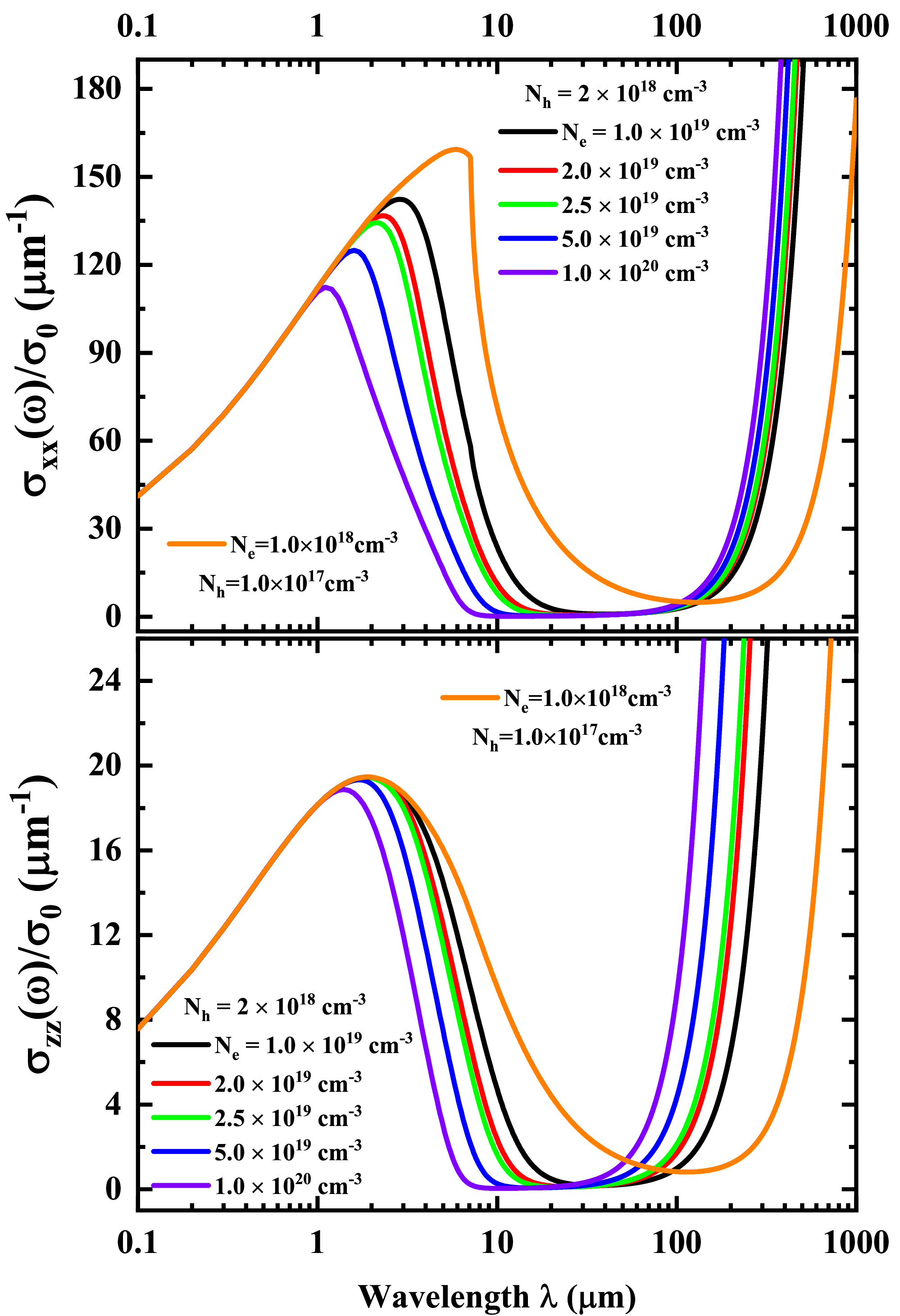}
  \caption{Optical conductivity spectra $\sigma_{xx}$ (upper panel) $\sigma_{zz}$ (lower panel)
  at fixed temperature $T=150$ K, electronic relaxation time $\tau=6.71$ ps, hole density $N_h=2.0\times10^{18}$ cm$^{-3}$ for different
  electron densities: $N_e=1.0\times10^{19}$ cm$^{-3}$ (black curve), $2.0\times10^{19}$
  cm$^{-3}$ (red), $2.5\times10^{19}$ cm$^{-3}$ (green), $5.0\times10^{19}$
  cm$^{-3}$ (blue), $1.0\times10^{20}$ cm$^{-3}$ (purple). We also show the corresponding results for a low carrier
  density sample $N_e=1.0\times10^{18}$ cm$^{-3}$ and $N_h=1.0\times10^{17}$ cm$^{-3}$ (orange curve).}\label{fig7}
\end{figure}

In the calculations we take the electronic energy relaxation time $\tau$ as an input
parameter that takes into account the effect of the broadening of the scattering states.
The dependence of $\sigma_{xx}(\omega)$ and $\sigma_{zz}(\omega)$ on $\tau$ is shown in
Fig. \ref{fig6} for a fixed temperature and fixed electron and hole densities. A longer energy
relaxation time corresponds to a smaller broadening of the scattering state and,
thus, to a sample with higher carrier mobility. We note that in the short-wavelength
regime, both $\sigma_{xx}(\omega)$ and $\sigma_{zz}(\omega)$ do not depend on the energy
relaxation time $\tau$ and the corresponding optical absorption peaks do not change with varying
$\tau$. Thus, $\tau$ mainly affects the intra-band electronic transition. As shown in
Fig. \ref{fig6}, both $\sigma_{xx}(\omega)$ and $\sigma_{zz}(\omega)$ in the long-wavelength
regime show red shifts with increasing $\tau$. Hence, a wider window in the absorption spectrum
can be observed for a larger $\tau$ or for a Na$_3$Bi sample with a larger electron mobility.

In Fig. \ref{fig7}, we show optical conductivities $\sigma_{xx}(\omega)$ and $\sigma_{zz}(\omega)$
as a function of radiation wavelength $\lambda$ at fixed temperature, electronic relaxation
time and hole density for different electron densities $N_e$. In the short-wavelength regime,
both $\sigma_{xx}(\omega)$ and $\sigma_{zz}(\omega)$ do not depend on the electron density.
Blue shifts of the optical absorption peak and of the window can be found with increasing
electron density. This is because $E_F^e$ increases with $N_e$ so that the inter- and intra-band
electronic transitions require larger photon energies for electrons going from lower occupied
states to higher empty states by the absorption of the photons. The blue shift of the optical
absorption peak can be more obviously observed for $\sigma_{xx}(\omega)$. In Fig. \ref{fig7},
we also show the case with the carrier densities $N_e=1.0\times10^{18}$ cm$^{-3}$ and
$N_h=1.0\times10^{17}$ cm$^{-3}$. In this case, $E_F^e$ is inside the conduction band Berry
curvature regime. We find that at fixed temperature, the high-frequency edge of the optical
absorption window is sharper when $E_F^e$ is above the Berry curvature regime than when $E_F^e$
is inside the Berry curvature regime. The degree of sharpness for the low-frequency edge of
the optical absorption window depends very weakly on the electron density. This suggests
that the presence of the Berry curvature affects mainly the high-frequency edge of the
optical absorption window, i.e., the inter-band electronic transitions. From Fig. \ref{fig7},
we notice that the position of the optical absorption peak and window depend sensitively on
the electron density. From graphene we know that the inter-band optical absorption channel
can be switched on and off when $E_F^e$ is larger or smaller than $\hbar\omega/2$, which
is a consequence of the linear like electron energy spectrum in graphene \cite{Dong09}.
The results shown in Fig. \ref{fig7} indicate that akin to graphene-based 2D Dirac
electronic systems, the inter-band optical absorption can also be turned on and off by
varying the Fermi energy or electron density in Na$_3$Bi based 3DDESs by, e.g., applying a
gate voltage. Thus, this material can also be used as electro-optical modulators.
Furthermore, it should be mentioned that graphene based electro-optical modulators work
in the mid-infrared regime $\lambda\sim$ 10 $\mu m$ \cite{Hau04,Mo01,Dong09}. Na$_3$Bi
based electro-optical modulator should also work in the mid-infrared bandwidth
$\lambda\sim$ 10 $\mu m$ according to the results obtained from this study.

The results shown in Figs. \ref{fig4}-\ref{fig7} indicate that in the short-wavelength
regime where inter-band electronic transitions dominate, the optical conductivities
$\sigma_{xx}(\omega)$ and $\sigma_{zz}(\omega)$ for Na$_3$Bi based 3DDES do not depend on
temperature, electron density, and electronic relaxation time, similar to the case of
graphene. However in our case, $\sigma_{xx}(\omega)$ and $\sigma_{zz}(\omega)$ depend
sensitively on radiation frequency along with the appearance of the optical absorption peak.
This is the main difference of the optical conductivity between graphene and Na$_3$Bi 3DDES.
Moreover, because Na$_3$Bi based 3DDES is a bulk material with different electronic energy
spectra along different crystal directions, the optical conductivity $\sigma_{xx}(\omega)$
is always larger than $\sigma_{zz}(\omega)$ regardless of temperature, electron density
and relaxation time. This implies that a stronger effect on the optical absorption can
be achieved by the radiation field polarized along the 2D plane of Na$_3$Bi.

\section{\label{sec:leve4}{Conclusions}}
In this study, we have calculated the optical conductivity for Na$_3$Bi based three-dimensional
Dirac electronic system (3DDES) by using a simplified ${\bf k}\cdot{\bf p}$ model and
the energy-balance equation approach based on the semi-classical Boltzmann equation.
The effect of light polarization along different crystal directions has been taken
into account. For n-type 3DDESs, we have examined the effect of the Berry curvature in the
conduction band on the optical conductivity, i.e., optical absorption. Furthermore,
we have compared the results with those from a 2DDES such as graphene. The main
conclusions drawn from this study are summarized as follows.

For short radiation wavelength radiation ($\lambda< \sim 2$ $\mu m$), the optical
conductivities $\sigma_{xx}(\omega)$ and $\sigma_{zz}(\omega)$ are induced mainly
through inter-band electronic transitions accompanied by the absorption of a photon.
In this regime, $\sigma_{xx}(\omega)$ and $\sigma_{zz}(\omega)$ depend on the radiation
frequency and an optical absorption peak can be observed. This is the main difference
from the high-frequency optical conductivity for a 2DDES such as graphene where
$\sigma(\omega)=e^2/4\hbar$ is universal, and this difference is a consequence of the
3D nature of the electronic band structure of Na$_3$Bi single crystal. However, both
$\sigma_{xx}(\omega)$ and $\sigma_{zz}(\omega)$ do not depend on temperature and sample
parameters such as electron density and electronic relaxation time, similar to the case
of graphene. In the intermediate radiation wavelength regime ($5 \mu m<\lambda< \sim 200 \mu m$),
a strong effect of the absorption window can be observed in both $\sigma_{xx}(\omega)$
and $\sigma_{zz}(\omega)$. Similar to graphene, this window is induced by the different
energies that are required for the intra- and inter-band electronic
transition channels. Therefore, the width, height, and position of the optical absorption
window depend sensitively on temperature, electron density, and electronic relaxation time.
We find that the presence of the Berry curvature in the conduction band affects mainly the
high-frequency edge of the optical absorption window. In the long radiation wavelength
regime ($\lambda> \sim 100$ $\mu m$), the intra-band electronic transitions contribute
mainly to the optical conductivity, where both $\sigma_{xx}(\omega)$ and $\sigma_{zz}(\omega)$
increase sharply with radiation wavelength. This is similar to the dependence of the
optical conductivity on the radiation wavelength of the Durde optical conductivity for
free electrons. Moreover, we have found that in a Na$_3$Bi based 3DDES, $\sigma_{xx}(\omega)$
is about five times larger than $\sigma_{zz}(\omega)$ in short-wavelength regime, regardless of temperature, electron
density and relaxation time. This effect is a consequence of the crystal structure of
Na$_3$Bi in which the $xy$-plane is a layered structure stacked along the
$z$-axis. Thus, a larger optical conductivity or absorption can be achieved by radiation
field polarized along the 2D plane of Na$_3$Bi.

From the viewpoint of device applications, the most significant theoretical finding from
this study is that akin to graphene-based 2DDES, the inter-band optical absorption channel
can also be turned on and off by varying the Fermi energy or electron density in
Na$_3$Bi based 3DDESs by, e.g., applying a gate voltage. Hence, Na$_3$Bi based 3DDESs can
also be used as electro-optical modulators working in the mid-infrared bandwidth
$\lambda\sim$ 10 $\mu m$. We hope that our interesting theoretical results presented and
discussed here can be verified experimentally and can lead to the application of 3DDESs
in advanced optoelectronic materials and devices.

\section*{Acknowledgements}

This work was supported by the National Natural Science foundation of China (U1930116,
U1832153, 11764045, 11574319, 11847054) and the Center of Science and Technology of Hefei
Academy of Science (2016FXZY002). Applied Basic Research Foundation of Department
of Science and Technology of Yunnan Province (No. 2019FD134), the Department of Education
of Yunnan Province (No. 2018JS010), the Young Backbone Teachers Training Program of Yunnan
University, and by the Department of Science and Technology of Yunnan Province.

\end{document}